# Mechanistic Insights into the Early Stages of Oxidation at Copper Terrace: The Role of O-O Repulsion and Substrate-mediated Effects

E V Charan Reddy and Abhijit Chatterjee*

*Department of Chemical Engineering, Indian Institute of Technology, Bombay, India*

*Corresponding author email: abhijit@che.iitb.ac.in

**Abstract**

Copper-based catalysts play a crucial role in industrial oxidation reactions. Although many theoretical studies consider copper to be metallic, it is well established that copper readily oxides at ambient conditions, forming a passivating oxide layer. Experimental investigations spanning two decades have shown that in addition to the anticipated step-oxide formation, oxide can directly form at the Cu(111) terrace. The atomistically-resolved mechanism for direct oxidation at flat terraces remains unknown. Using density functional theory (DFT) calculations, we demonstrate that the formation of subsurface oxide occurs through a coordinated mechanism that takes place in the presence of specific clusters of adsorbed oxygen atoms. Certain oxygen atoms in the cluster function like pincers to extract a copper atom from the surface layer and induce localized surface restructuring. This process creates open channels that allow an oxygen atom to diffuse into the subsurface layer. The subsurface oxide formation is barrierless. This implies that the Cu oxide surface is highly dynamic. At low O coverages, subsurface oxidation is unlikely via step oxide growth nor direct terrace oxidation as the subsurface oxygen is unstable. Substrate mediated O-Cu-O adsorbate interactions govern the oxide stability. These insights provide a foundation for developing a more accurate dynamic models for copper catalysis.

**Table of contents image**

## 1. Introduction

Copper is a catalyst of significant industrial importance. It plays a crucial role in various applications, such as low-temperature CO oxidation[1], ethylene oxidation[2] and the methanol steam reforming[3,4]. Adsorbed oxygen (O*) is a key reaction intermediate in these processes. O* can also induce the oxidation of copper under ambient conditions, initially leading to the formation of a thin $Cu_2O$ layer[5]. The copper oxide layer profoundly influences catalytic activity, necessitating a mechanistic understanding of the early stages of oxide formation[6].

Experimental literature on the onset of oxide formation spans more than two decades[5,6,15–19,7–14]. Matsumoto *et. al*[12], and others[6,13] employed scanning tunnelling microscopy (STM) and low-energy electron diffraction (LEED) at room temperature to study the process. Cu oxidation is facile. Very low oxygen pressures, for instance, $10^{-8}$ to $10^{-5}$ Torr, are maintained to ensure that the oxygen coverage remains low. Oxidation of an initially clean Cu(111) surface is initiated rapidly at the step edges (termed "step oxide") which propagates to the upper terraces. Oxide can directly form even at flat terrace regions. They reported triangular depressions observed in STM images and termed these regions as defect sites that lead to the "terrace oxide". As the O exposure increases, similar features also emerge on defect-free terraces ("added oxide"), which dominate the surface in later stages and disrupt terrace oxide growth. Recent study with time-lapse NAP-XPS measurements showed that both oxidation processes began with adsorbed oxygen atoms and progressed to the formation of the $Cu_2O$ phase[17]. Although these observations are intriguing, experimental techniques are unable to access short timescales, such as, nanoseconds, which is needed to build a comprehensive dynamical atomic scale understanding.

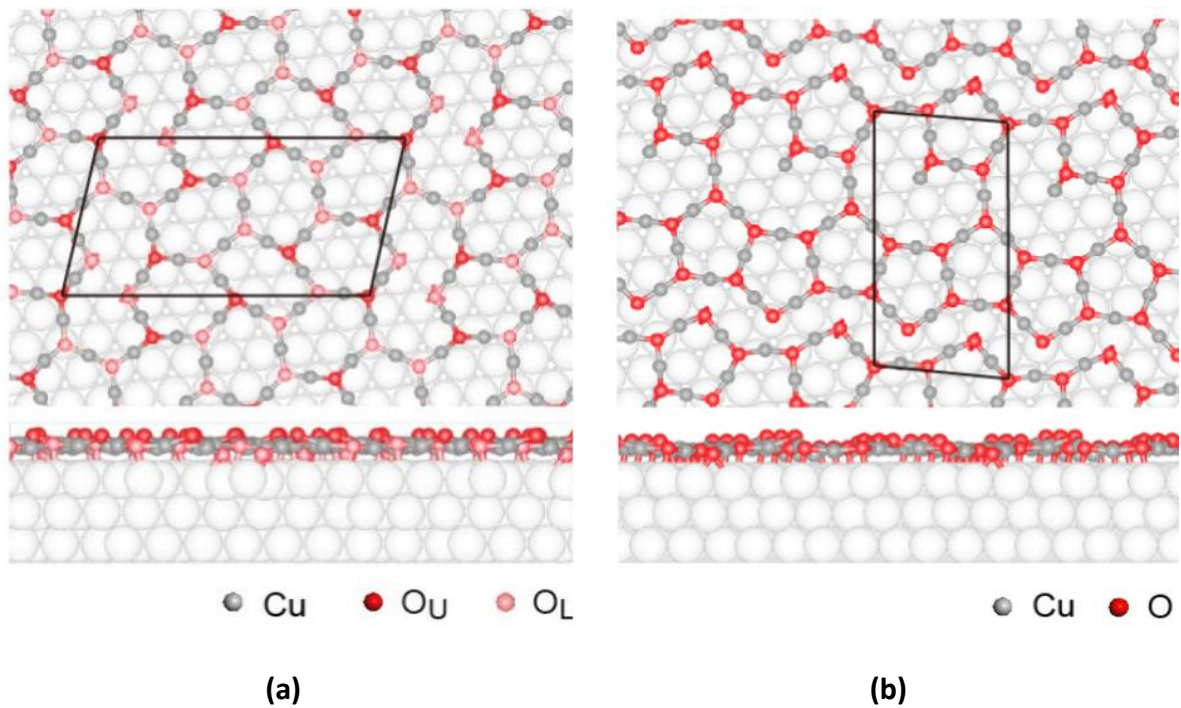

(a)            (b)

**Figure 1: Structural model[19] of a) "44" type oxide layer b) "29" type oxide layer on top of Cu(111) surface. Taken from Ref. [19]. ($O_U$ is the upper layer O atom and $O_L$ is the lower layer O atom.)**

The initial oxidation of metal surfaces is believed to progress through a sequence of stages: oxygen chemisorption, subsurface incorporation, and lattice transformation into an oxide phase[20]. Above 170 K, $O_2$ dissociates into atomic oxygen (O) at low exposure[18,21], preferentially occupying three-fold hollow (fcc/hcp) sites[8]. Elevated exposures induce disordered oxygen overlayers, triggering Cu adatom mobility and metastable reconstructions[7]. These transient structures serve as precursors to surface oxide formation, ultimately evolving into ordered $Cu_2O$ layers[12]. Although the exact mechanism is unclear, combined experimental and computational studies were able to identify two stable $Cu_2O$-like superstructures, termed "29" and "44" oxides, when the Cu(111) surface was annealed at 423 and 673 K, respectively[9,10,12,16] (as shown in Fig. 1). At temperatures exceeding 300 K, these metastable phases evolve into well-defined thin oxide films[12]. Subsequent oxide growth requires substantial metal atom incorporation at the metal-oxide interface, driven by either bulk metal-atom diffusion to the surface or subsurface oxygen migration[22].

To further underscore the need for a deeper atomic scale understanding, Matsumoto et. al[12] reported the ability of oxygen to extract Cu atoms directly from pristine terraces during room-temperature Cu oxidation without intrinsic defect, which they termed as "added oxide" [12]. Unlike step-edge-mediated oxidation[14], this process involves oxygen-induced destabilization of the terrace lattice, enabling Cu adatom extraction and oxide nucleation. The atomistic mechanism driving this behaviour, such which Cu sites are prone and the conditions, remains unresolved. The main objective of this study is to understand the conditions in which subsurface oxide can form at the defect free terrace region and reveal the oxidation mechanism in the process.

Since substrate-mediated O–Cu–O interactions are expected to play a central role in the terrace oxidation, we employ density functional theory (DFT) calculations to explore such many-body interactions. Our calculations show that indeed clustered oxygen adsorbates induce localized surface restructuring, weakening the Cu–Cu bonds and displacing Cu atoms. We show that this facilitates oxygen ingress into the subsurface layers. Nudged elastic band (NEB) calculations reveal that subsurface oxide formation becomes barrierless once O–Cu–O configurations stabilize, aligning with operando observations of rapid terrace pit formation.

## 2. Methodology

All plane-wave density functional theory (DFT) calculations were performed using the Vienna Ab initio Simulation Package (VASP) version 5.4.4[23–25]. The electronic convergence tolerance was set to $10^{-5}$ eV with a force tolerance of 0.01 eV/Å. The exchange-correlation energy was described using the self-consistent Generalized Gradient Approximation (GGA) based Perdew, Burke, and Ernzerhof (PBE) functional[26]. The plane-wave basis set utilized an energy cutoff of 450 eV, while the interaction between core and valence electrons was modelled using the Projected Augmented Wave (PAW) potentials[27].

For the Cu lattice point optimization calculations, the Methfessel-Paxton electron smearing was employed (with a smearing width of 0.1 eV), and the Brillouin zone was sampled with a 16×16×16 k-point mesh. The optimized Cu lattice constant was found to be 3.635 Å (which is in reasonable agreement with the experimentally known Cu lattice constant[28] 3.615 Å).

The Cu(111) surface was modelled using a three-layer (4×4) slab with a vacuum region of 12 Å (as shown in Fig. 2) to accommodate adsorbed O species while minimizing interactions with periodic images. To simulate surface adsorption, the bottom layer was kept frozen, while the top two layers are allowed to relax until the forces were less than 0.01 eV/Å. The Brillouin zone sampling was performed by using a 6×6×1 Monkhorst-Pack k-point mesh. The binding energy of an adsorbate species (X) is defined by

$$E_{ads}(s) = \frac{E_{n_X-Cu} - E_{Cu} - n_X E_X}{n_X}. \qquad (1)$$

Here, $n_X$ is the number of adsorbed O atoms, $E_{Cu}$ is energy of the bare copper surface (reported in units of eV/O-atom), $E_{n_X-Cu}$ is the energy of the configuration consisting of $n_X$ adsorbed atoms and $E_X$ is the energy of an isolated oxygen atom in the gas phase.

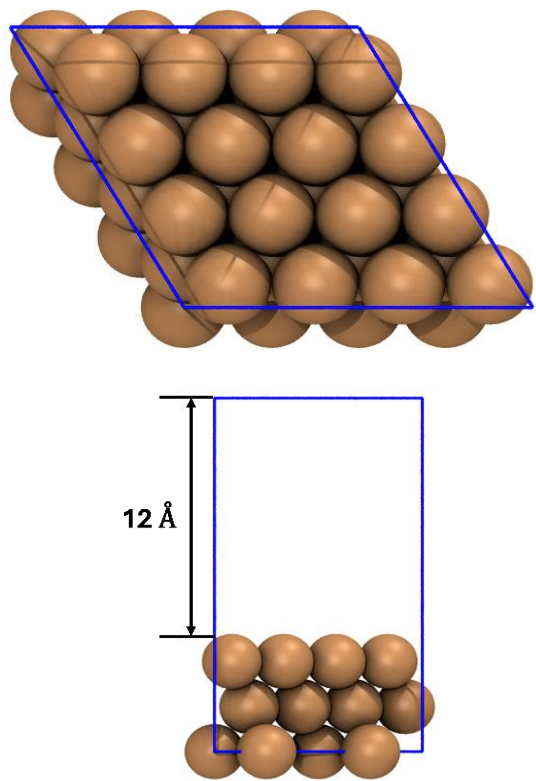

**Figure 2:** Top and side views of the Cu(111) slab used in the simulations.

The minimum energy pathways and corresponding energy barriers for subsurface oxygen diffusion were computed using the nudged elastic band (NEB) method[29]. Each diffusion pathway was discretized with a minimum of six intermediate images, including the initial and final configurations. Transition states were identified and verified through vibrational frequency analysis, which confirmed the presence of a single imaginary frequency along the reaction coordinate.

## 3. Results and Discussion

### 3.1. Single O atom binding on Cu(111) surface

**Table 1: Binding energies of an O atom on Cu(111) surface.**

| Adsorbed species | Site | Binding energy (eV) |
|---|---|---|
| Oxygen (O*) | fcc | -5.000 |
| | hcp | -4.905 |

The binding of a single oxygen (O*) at different sites of the Cu(111) surface were evaluated. It is found that threefold hollow sites, both fcc and hcp, as shown in Fig. 3, are the energetically preferred binding sites for O* with a binding energy of -5 eV and -4.905 eV, respectively. was calculated and is presented in Table 1. Compared to the fcc site, O binds slightly weaker at the hcp site. The predicted stable sites agree well with the experimental findings[8,11] and the calculated binding energies are in reasonable agreement with the experimental values (-4.46[30], -4.60[31], -4.77[32] eV). The calculated Cu-O bond distances at the fcc and hcp sites are 1.06 and 1.099 Å, respectively, as shown in Fig. 3. No corrugations at the Cu surface are observed.

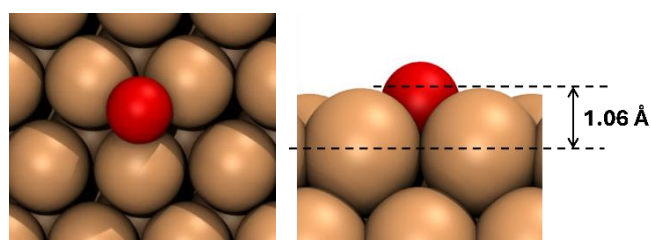

**O* at FCC site**

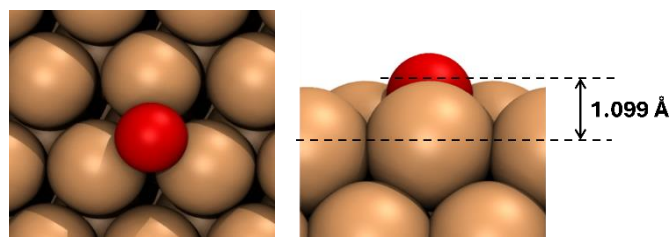

O* at HCP site

**Figure 3: Side (right) and top (left) views of the most stable binding sites of a single O* atom on Cu(111) surface.**

The initial stages of Cu(111) oxidation are marked by the formation of disordered chemisorbed oxygen overlayers. These overlayers, whether ordered or disordered, consist of oxygen adatom clusters whose mutual interactions govern the structural evolution of the surface[33]. To elucidate the thermodynamic stability and interaction mechanisms underlying these disordered phases, we present a comprehensive investigation of oxygen adatom clusters-including pairs, triplets, and extended aggregates-across diverse atomic configurations. This many-body interaction term (in units of eV/O-atom) is evaluated as[34]

$$E_{Many-body} = E_{ads}(s) - E_{ads}^0. \qquad (2)$$

Here, $E_{ads}(s)$ is the binding energy (eV/O-atom) of a configuration that contains more than one adsorbed O* atom and $E_{ads}^0$ is the binding energy for an O* atom without any O* neighbours. $E_{Many-body}$ is zero when the adsorbed O* are far apart. As the O* atoms are brought together there is significant lateral repulsion among the adsorbed species, resulting in $E_{Many-body}$ becoming positive. This repulsion dictates the morphology of the adsorbate overlayer. Difference in $E_{Many-body}$ for two different configurations with the same value of $n_X$ is related to the relative probability of the configurations. Furthermore, thermodynamically preferred configurations with high values of $E_{Many-body}$ can be accessed at large oxygen chemical potentials. Configurations were prepared by fixing one oxygen atom at a stable surface site (fcc or hcp) and systematically varying the positions of additional O* atoms across nearest-neighbour (NN) sites. We discuss only those O* arrangements where $E_{\text{Many-body}}$ exceeds 20 meV/O-atom. Fig. 3 is applicable configurations involving weaker interactions.

## 3.2. Oxygen pair cluster

For pair clusters, we constructed configurations by fixing one O* atom at an fcc or hcp site, and then systematically placing the second O* atom at first-, second-, and third nearest-neighbour (NN) sites. The first-NN hcp position resulted in significant O–O repulsion, causing it to be energetically unstable.

Fig. 4 plots the two-body O*–O* interaction energies as a function of pair distance. The interactions are positive valued, revealing their repulsive nature. The many-body interaction energy for the 2-NN O–O fcc site is 0.198 eV and decays with increasing O–O separation, reaching 0.014 eV at the 6-NN hcp position—beyond which interactions become negligible.

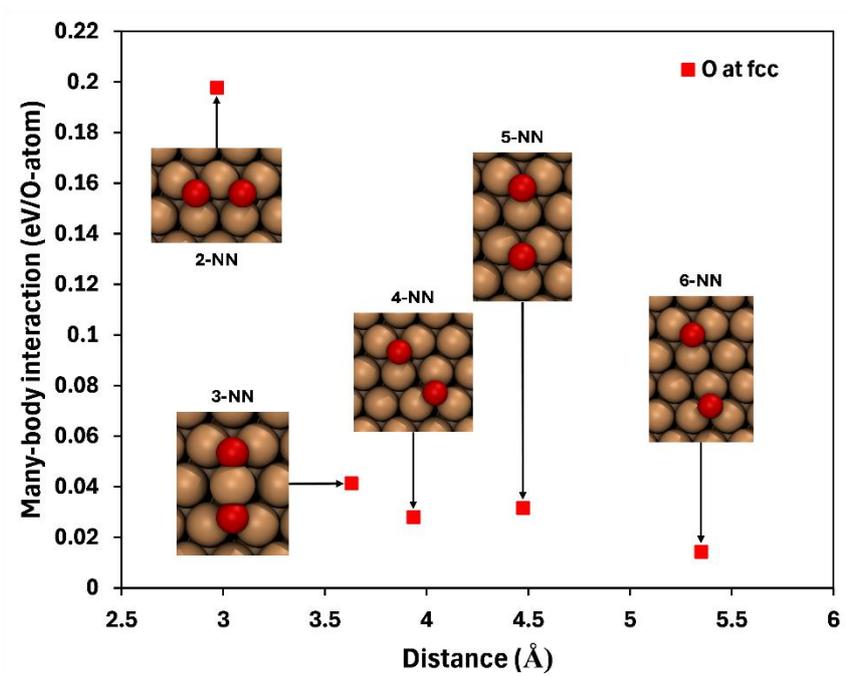

(a)

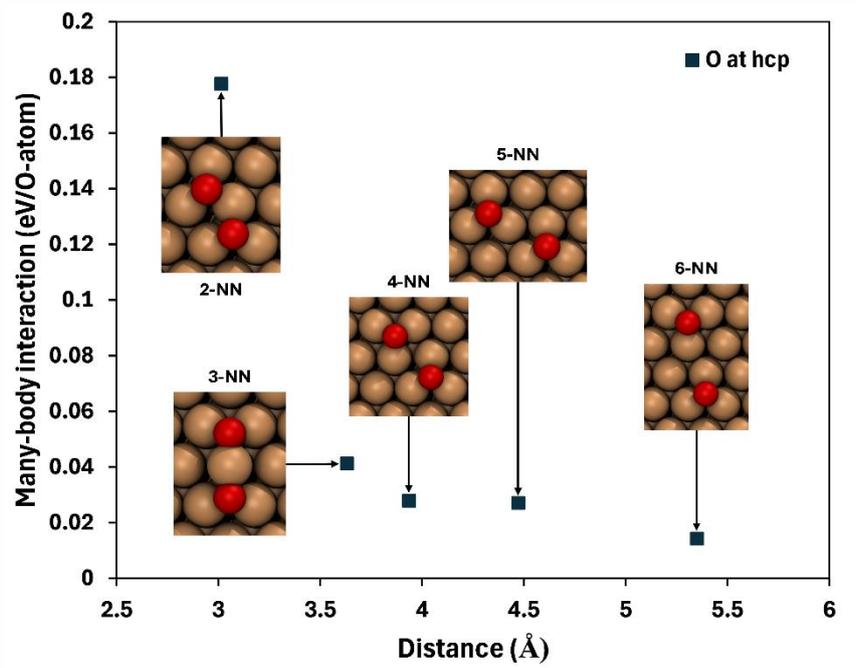

(b)

Figure 4: Many-body interaction term vs. distance between an O*–O* pair. First O* atom is placed at (a) an fcc site, and (b) hcp site. The energy minimized configurations are also shown.

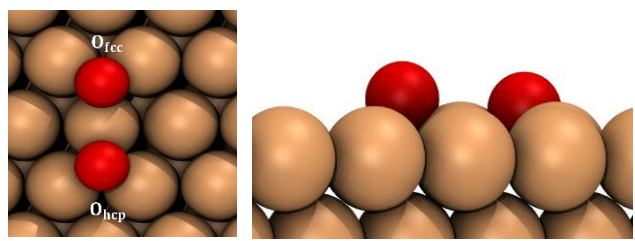

**(a) Input structure**

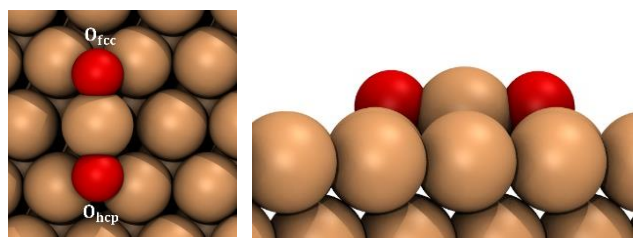

**(b) Optimized structure**

**Figure 5: Side (right) and top (left) views of the (a) input structure for DFT calculation (b) optimized structure after DFT calculation where one O atom is fixed at fcc position and other O is at 3-NN hcp position.**

A striking feature observed is the sharp reduction in the interaction from the 2-NN to 3-NN and farther positions (Fig. 3). Figure 4 reveals the effect of the strong substrate-mediated oxygen-oxygen interactions. The initial configuration where a pair of O-O atoms at 3-NN separation is shown in panel a, whereas the DFT-optimized configuration is shown in panel b. It is observed that the O* atoms, one present at fcc site and the other at hcp site, trigger substantial copper lattice reconstruction. Notably, the copper atom undergoes displacement of 0.978 Å from its equilibrium position—comparable to the Cu–O bond length itself. The oxygen pair effectively functions as a pincer, extracting the copper atom (compare panels b and c to better appreciate the surface corrugation). The O–O repulsion is mediated by the Cu substrate. This structural rearrangement, combined with the increased O–O separation, results in a significant reduction in the many-body interaction energy. In fact, the many-body interaction at 3NN distance is comparable to 4NN distance, i.e., ~0.04 eV/O atom. The pronounced surface corrugation arises from the substantial electronegativity difference between O and Cu (~1.54[35]) . The effects of oxygen adsorption are primarily localized on the oxygen atom and its nearest-neighbour substrate atoms. Due to its high electronegativity, the oxygen atom withdraws electron density from the 1-NN copper atoms, inducing an inward dipole moment that drives the copper atom to deviate from its bulk position[36]. The pincer-like behaviour is not seen in case of 2-NN separation between the O-O pair.

From Fig. 5, we find that even at low coverage of ⅛ ML substrate-mediated interactions can exhibit pronounced influence. With increasing coverage, these interactions might drive significant reorganization of adsorbate configurations. To elucidate the emergent cooperativity within multi-adsorbate ensembles, we conducted systematic DFT simulations across higher-order clusters—including triplets, quadruplets, quintets, and sextets to investigate coverage-dependent evolution of such multi-adsorbate ensembles.

### 3.3. Triplet and higher order cluster types

At higher oxygen chemical potentials, O-pairs can be present at second neighbour separation as already seen in Fig. 5. We focus on O-triplets again at the same separation as shown in Fig. 6. In such a case, the O atoms occupy either the fcc or hcp sites () and forming an equilateral triangle. For this equilateral triangle configuration, O atoms can arrange themselves in two ways which are shown in Fig. 6 (a) and 6 (c). For simplicity, the configuration in Fig. 6(a) is called triplet-1, and the configuration in Fig. 6(c) is called triplet-2. An interesting aspect about these O* arrangements is that despite being geometric mirror images, triplet-1 and triplet-2 are topologically different: triplet-1 lacks a Cu atom at the centroid of the triangle, while

triplet-2 features a Cu atom at the centre which is marked in red colour in the Fig. 6(c). The DFT optimized configurations are shown in Fig. 6 (b) and 5 (d).

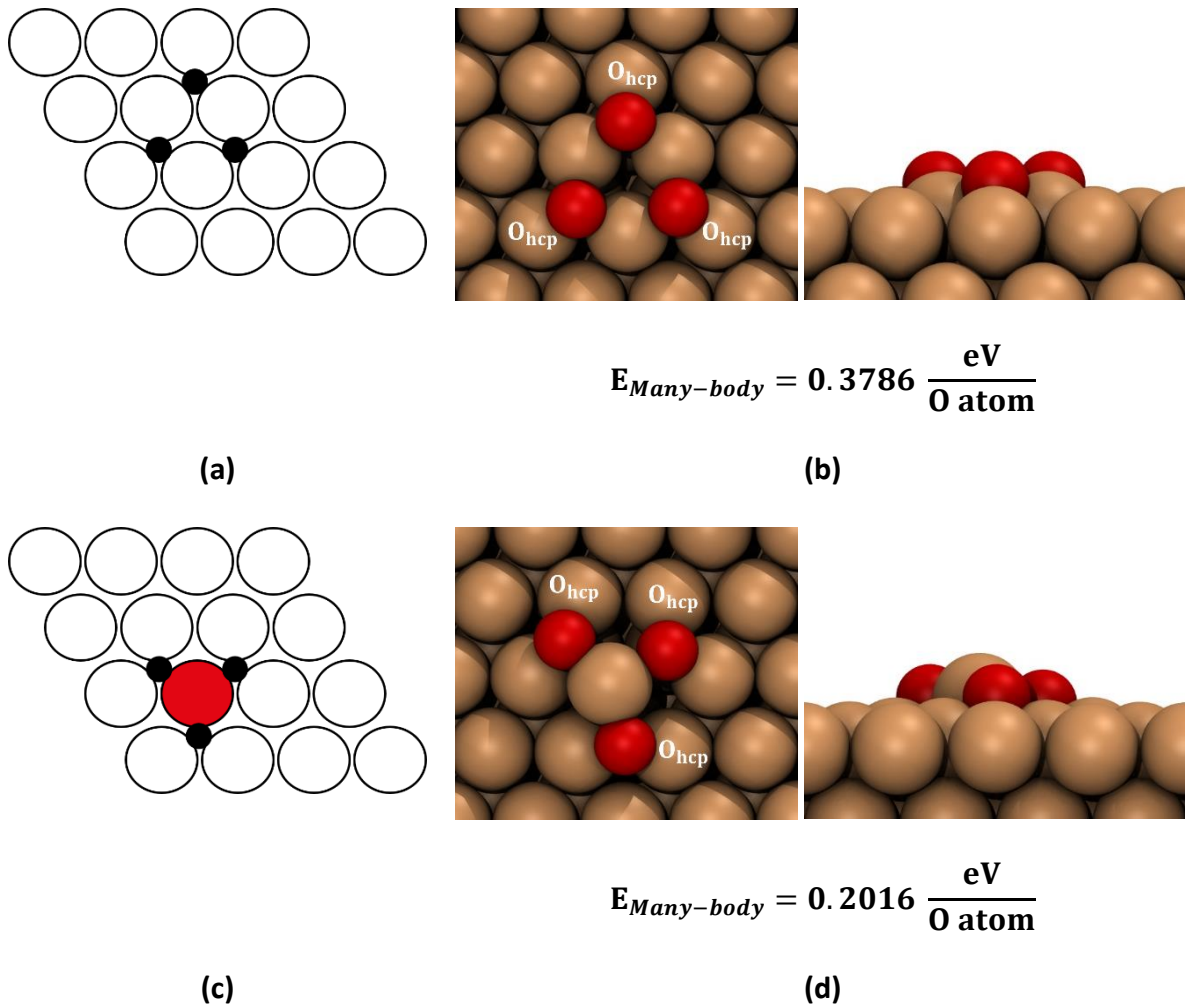

Figure 6: (a), (c) input structure for DFT calculation; side (right) and top (left) views of the (b), (d) optimized structure after DFT calculation where O atoms are occupied at hcp position in a triplet arrangement.

A common assumption made in the cluster interaction modelling community is that many-body interactions depend only on the adsorbate arrangement, and that symmetrically equivalent adsorbate arrangements are associated with the same interaction energy independent of substrate site symmetry. Despite being geometric mirror images, triplet-1 has a many-body interaction of 0.379 eV/O-atom and triplet-2 has a many-body interaction of 0.202 eV/O-atom. This significant disparity in many-body interaction (~0.177 eV/O-atom) is due to the copper atom present at the centroid.

In triplet-2, the central Cu atom experiences strong attractive forces from three adjacent O atoms, inducing a vertical displacement of the Cu atom by 1.286 Å— which exceeds the Cu–O bond length. This displacement effectively transforms the Cu atom into an adatom, resulting in a O-Cu-O quadruplet cluster after electronic relaxation. Within this cluster, O–O repulsion is counterbalanced by O–Cu attraction, stabilizing the structure. This explains the relative stability of the triplet-2 arrangement over triplet-1. Interestingly, the many-body interaction for triplet-2 is nearly equal to the O–O pair interaction at 2-NN position as shown in Fig. 4(b).

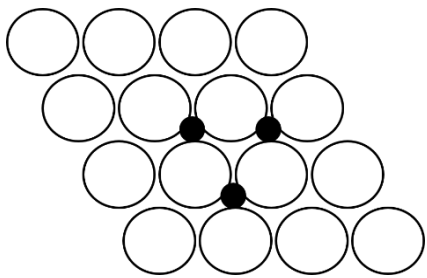
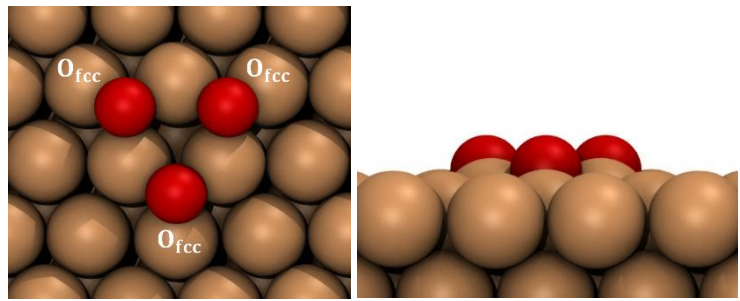

$$E_{Many-body} = 0.4097 \frac{eV}{O-atom}$$

(a)           (b)

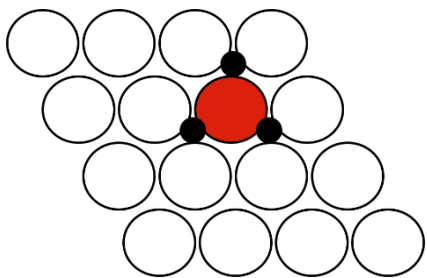
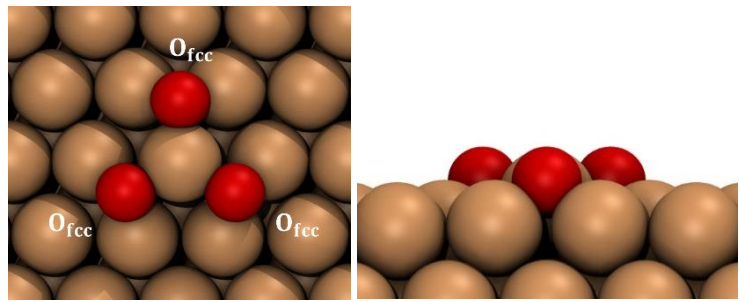

$$E_{Many-body} = 0.2842 \frac{eV}{O-atom}$$

(c)           (d)

**Figure 7: (a, c) input structure for DFT calculation; (b, d) side (right) and top (left) views of the optimized structure after DFT calculation where all O atoms are occupied at fcc position in a triplet arrangement.**

We further simulated triangular configurations with oxygen atoms occupying 2-NN fcc sites (Fig. 7 (a) and (c)), retaining the nomenclature used with $O_{hcp}$ triplets. Density functional theory (DFT) results for these configurations are presented in Fig. 7(b) and (d). A similar trend is observed for fcc-site configurations. The triplet-1 configuration exhibits a many-body interaction energy of 0.4097 eV/O-atom, while triplet-2 shows a lower value of 0.2842 eV/O-atom. The higher repulsion in fcc triplet compared to the hcp triplet arises from reduced substrate corrugation in the former. The central Cu atoms in the fcc triplet-2 undergo smaller displacements (~0.868 Å) compared to their hcp counterpart, which again highlights the role of substrate-mediated interactions. Moreover, it is interesting to note that site preference has flipped at higher O coverage. Now, the hcp sites become energetically more preferred over the fcc sites (see Table 1).

We extend our simulations to quadruplet O* clusters (at ¼ ML coverage). Figure 8 presents DFT results for fcc and hcp-based quadruplets. Configurations in Fig. 8 (a), (e) are designated as quadruplet-1, with their geometric counterparts termed as quadruplet-2 (Fig. 8(c), (g)). Analogous to triplets, quadruplets display distinct energetic and structural disparities which are shown in Fig. 8(b), (d), (f) and (h).

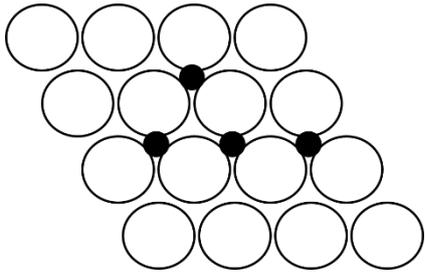
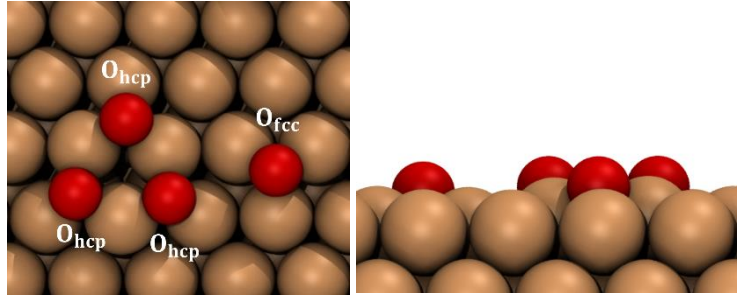

$$E_{Many-body} = 0.3235 \frac{eV}{O-atom}$$

**(a)**                        **(b)**

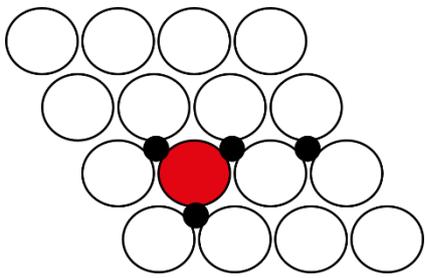
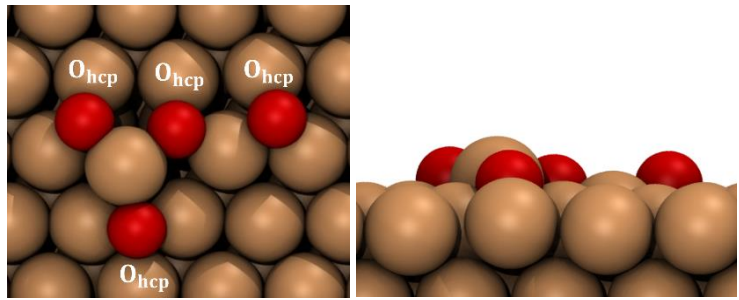

$$E_{Many-body} = 0.2665 \frac{eV}{O-atom}$$

**(c)**                        **(d)**

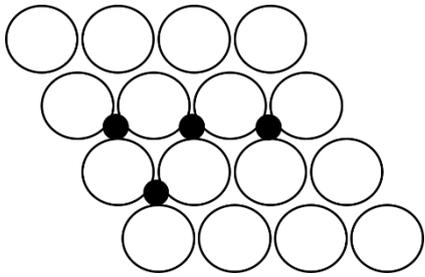
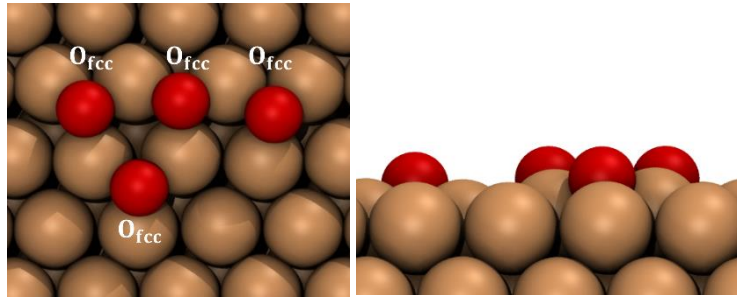

$$E_{Many-body} = 0.4829 \frac{eV}{O-atom}$$

**(e)**                        **(f)**

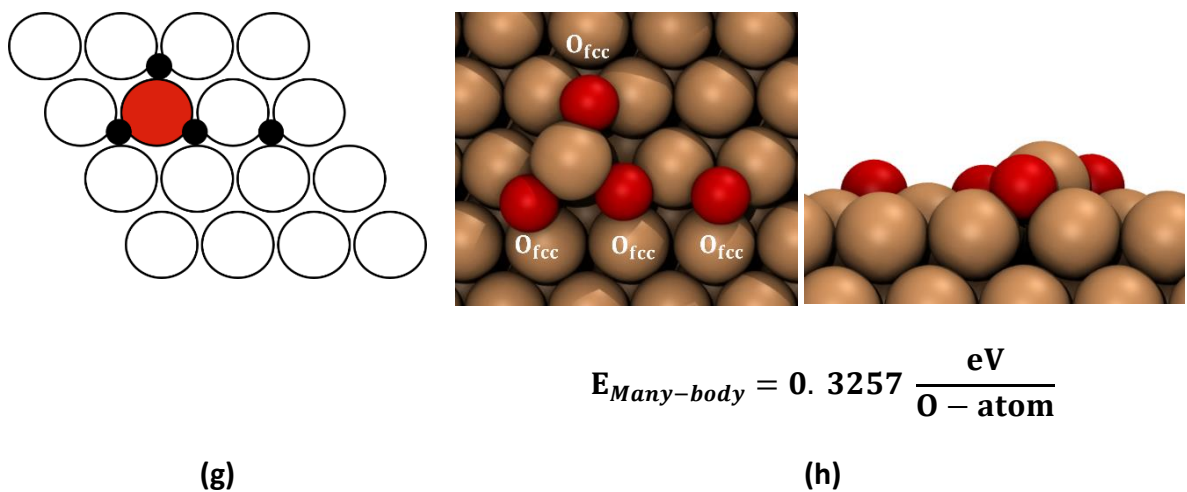

$$E_{Many-body} = 0.3257 \frac{eV}{O-atom}$$

(g)          (h)

**Figure 8: (a, c) Initial O-quadruplet structures with oxygen atoms at hcp sites, used as inputs for DFT calculations. (e, g) Initial O-quadruplet structures with oxygen atoms at fcc sites, used as inputs for DFT calculations. (b, d) Side (right) and atop (left) views of the DFT-optimized structures for hcp-site quadruplets. (f, h) Side (right) and atop (left) views of the DFT-optimized structures for fcc-site quadruplets.**

For hcp-based quadruplets, quadruplet-1 exhibits a many-body interaction energy of 0.3235 eV/O-atom, while quadruplet-2 shows 0.2665 eV/O-atom. In quadruplet-1, heightened O–O repulsion causes one O* atom to be displaced from an hcp to a neighboring fcc site (which can be seen by carefully comparing the initial and final configurations in Fig. 8 (a) and (b), respectively). For the fcc-based quadruplets, quadruplet-1 exhibits a many-body interaction energy of 0.483 eV/O-atom, whereas quadruplet-2 shows 0.326 eV/O-atom. In quadruplet-2 configuration, the central Cu atom is displaced by approximately ~1.47 Å —about 1.7 times greater than in the fcc triplet-2 configuration (which can be observed in Fig. 7 (d) and 8 (h)), a consequence of intensified O–O repulsion at elevated coverages.

Extending the simulations to higher-order clusters such as quintets, sextets (see Supplementary Information) reveals increasingly pronounced substrate distortion. For sextets, significant surface reconstruction occurs, marked by pronounced displacement of surface Cu atoms (~1.2 – 1.5 Å), signaling adsorbate-driven lattice instability. Across all clusters, the common theme is the stabilization of cluster-2 configurations over cluster-1 arises from a dynamic balance between substrate-mediated interactions and O–O repulsion. Notably, triplet-2 units serve as the foundational building blocks in the higher-order clusters, with configurations incorporating these units exhibiting prolonged Cu atom displacements from their lattice positions. This cooperative stabilization mechanism highlights the critical role of adsorbate-induced substrate restructuring in governing oxide overlayer evolution at high coverages.

Observations from Fig. 6(d), 8(d) and 8(h) reveal that the upward displacement of surface Cu atom is accompanied by the partial downward movement of O* atom(s) toward subsurface positions within clusters. This raises critical questions: Is subsurface oxygen diffusion feasible at high coverages and is the process dependent on the O* coverage?

### 3.4. Onset of subsurface oxidation

Our simulations reveal that a single oxygen atom in the subsurface octahedral site is thermodynamically unstable, as it spontaneously moves to the surface fcc site to become an O* atom (Supporting Information). This result contradicts a prior report of a stable subsurface oxygen at octahedral sites with a binding energy of ~3.2 eV[36], i.e., an energy difference of +1.8 eV between the subsurface and surface O states was reported. The discrepancy, in terms of stability, arises from differences in the convergence criterion used. The earlier study used weaker convergence thresholds (energy tolerance: $3\times10^{-3}$ eV; force tolerance: 0.07 eV/Å), which leads to a conclusion that the subsurface O configurations is stable, compared to a more stringent criteria used here (energy: $10^{-5}$ eV; force: 0.01 eV/Å). This confirms that, subsurface oxygen diffusion is non-spontaneous at very low coverages ($1/16$ ML). The instability arises from lattice strain caused by the incorporation of subsurface oxygen, which imposes a substantial energetic penalty. The Cu(111) surface resists such distortions to retain its thermodynamically favoured lattice configuration, rendering subsurface oxygen unstable. Therefore, the importance of the step edge oxidation mechanism is diminished. Even in situations where the subsurface oxygen were to arrive from the step regions, the fact that subsurface oxygen is inherently unstable is an important point in the overall context of O* coverage dependent oxidation of pristine Cu(111) surface.

To elucidate coverage-dependent stability of subsurface O, we extend our DFT simulations to higher-coverage configurations (triplets and quadruplets). In Fig. 9, we systematically optimized one of the oxygen atoms within the hcp triplet-2 by initially keeping it at the subsurface position, as shown in panel (a). However, the relaxed configuration in Fig. 9(b) demonstrates that subsurface oxygen remains unstable at 3/16 ML O* coverage, diffusing back to surface sites.

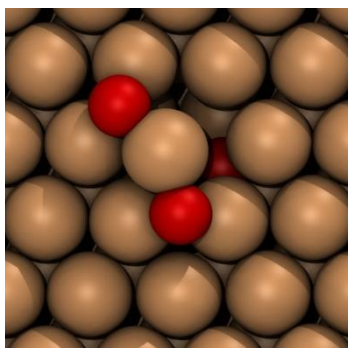   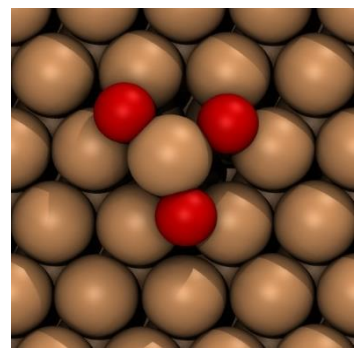

**(a)**            **(b)**

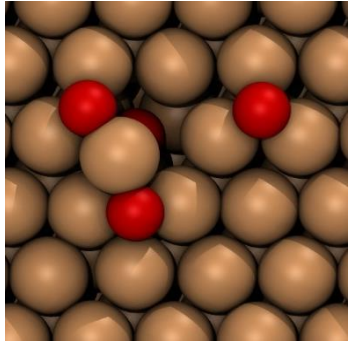 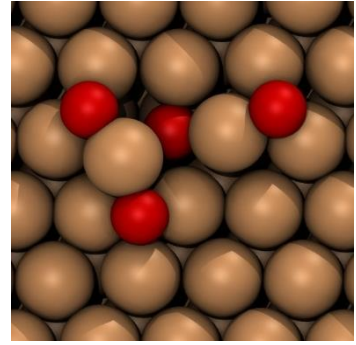

$$E_{ads} = -4.667 \frac{eV}{O-atom}$$

(c)          (d)

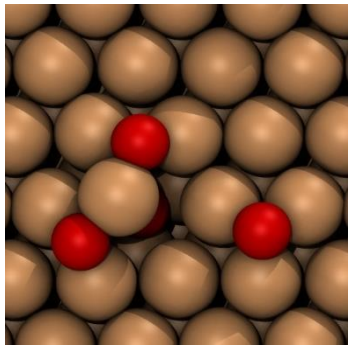 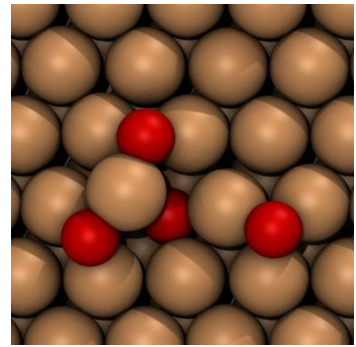

$$E_{ads} = -4.689 \frac{eV}{O-atom}$$

(e)          (f)

**Figure 9: (a) Initial configuration for DFT calculations, with the one of the oxygen atoms positioned at an octahedral site. (b) Optimized structure following DFT relaxation, showing that configuration in panel (a) is unstable, and that the oxygen atom prefers being at the surface hcp site. (c, e) Initial DFT configurations with one oxygen atom positioned at subsurface sites in hcp (a) and fcc (c) quadruplet-2 frameworks. (d, f) DFT-optimized structures reveal subsurface oxygen stabilization.**

On the other hand, quadruplet configurations, involving both fcc and hcp-based O* clusters, exhibit a propensity for subsurface diffusion (as seen in Fig. 7(d), 7(h)). To probe this further, we optimized one oxygen atom in each configuration at subsurface positions. DFT-optimized energetically stable configurations are presented in Figure 9. Figures 9(d) and 9(f) reveal that subsurface oxygen in relaxed hcp and fcc quadruplet configurations results in binding energies of −4.667 eV/O-atom and −4.689 eV/O-atom, which are closed to the nearly equivalent to the respective surface-adsorbed counterparts, i.e., xx and xx eV/O-atom, respectively (also see Section xx in Supplementary Information). This indicates thermodynamic stability of subsurface oxygen at elevated coverages (¼ ML). The stabilization again arises crucially from the interplay of substrate-mediated interactions and O–O repulsion. The cooperative O–Cu bonding induces localized Cu atom displacements (~0.98 Å) from lattice positions, creating structural distortions that accommodate subsurface oxygen. At higher coverages, repulsive interactions between adsorbed oxygen atoms amplify these distortions, effectively allowing them to act as pincers, and widening the open channels which enables subsurface incorporation of larger atoms such as oxygen on Cu(111).

To quantify the energy barrier for this process, we performed nudged elastic band (NEB) calculations, mapping the diffusion pathway and transition states of oxygen migration from surface to subsurface sites. NEB calculations were performed for hcp and fcc quadruplet configurations, using the initial and final states taken from Fig. 8(d), 8(h) and Fig. 9(d), 9(f) respectively. The results are shown in Figure 10. Snapshots of intermediate images are provided in the Supporting Information. Together these provide a mechanistic understanding and the conditions for the "added-oxide" process.

The energy barrier for subsurface oxygen diffusion in hcp quadruplets is 4.7 meV (Fig. 10(a)), while for fcc quadruplets, it is 19 meV (Fig. 10(b))— both are significantly lower than the thermal energy at room temperature ($k_BT \approx 25.7$ meV). These results indicate that subsurface oxygen diffusion is effectively barrierless at the elevated O* coverages. The adsorbed O* species can rapidly move between the surface and sub-surface position. The copper oxide surface is a highly dynamic Once again this underscores the importance of the "added oxide" mechanism and the diminished role of the step-oxide formation.

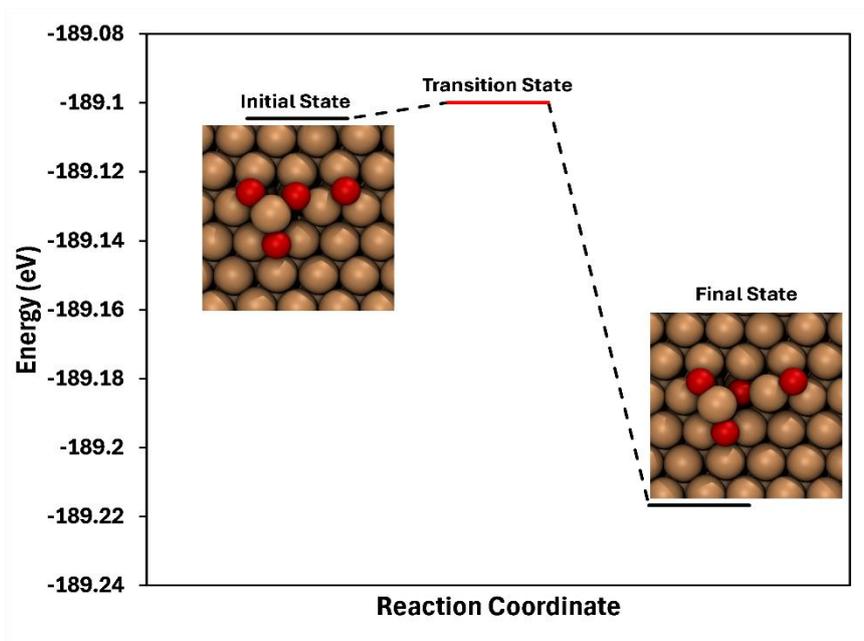

(a)

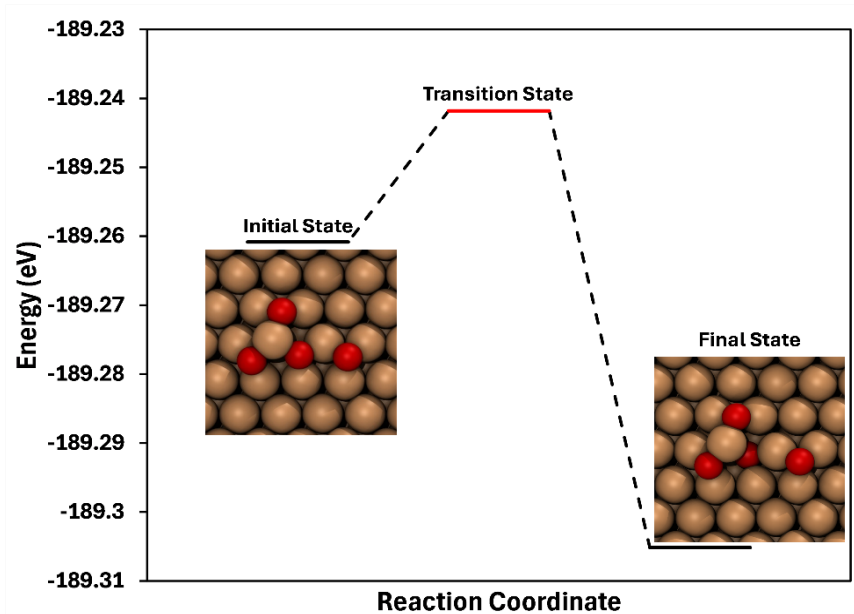

(b)

**Figure 10:** a) Surface-to-Subsurface oxygen diffusion pathway for hcp quadruplet-2. b) Surface-to-Subsurface oxygen diffusion pathway for fcc quadruplet-2 configuration.

## 4. Conclusions

The formation of surface oxide the defect-free Cu(111) terrace is of tremendous importance to heterogeneous catalysis. Our present work provides a comprehensive atomistic-scale understanding of the conditions at which the oxide can form and addresses few long-standing questions about the underlying mechanism[37]. The oxidation process is found to be sensitive to the oxygen coverage, the oxygen arrangement, and the substrate-mediated O-Cu-O interactions. When 3-oxygen atom and larger clusters occupy the hcp hollow sites, these clusters can destabilize the substrate lattice and induce localized surface reconstructions. It is interesting to note that the fcc sites which are preferred at low coverage, become less preferred at the higher coverage. The oxidation process is a cooperative phenomenon wherein the interplay between attractive O–Cu interactions and repulsive O-O interactions causes the surface to reconstruct and open up diffusion channels at the surface. The open channels enable facile back-and-forth oxygen movement between the surface and the subsurface layer, which is the critical step for nucleating oxide layers on flat terraces. The oxidation process is barrierless under such conditions. The Cu surface reconstruction is not witnessed especially at low oxygen coverage. As a result, the subsurface oxygen is found to be unstable at these coverages. In summary, we conclude that surface oxidation mechanism at the defect-free Cu(111) terrace is a highly dynamic process. which has major implications in heterogeneous catalysis.

To further highlight the importance of this study, it should be noted that oxide layer formation is a ubiquitous feature of catalytic processes such as CO oxidation and methanol synthesis on Cu(111) surfaces. Given that key intermediates (e.g., CO, O) drive adsorbate-induced surface reconstructions[38], the proposed cluster-mediated oxidation mechanism provides critical insights for modelling of these dynamic catalytic systems. By explicitly incorporating adsorbate-driven lattice distortions and subsurface diffusion pathways, this framework enables predictive simulations of transient surface states during reaction cycles—a prerequisite for designing catalysts with tailored activity and stability.

More broadly, the present work contributes to the large body of extensive experimental and computational studies that have established that strongly chemisorbed species drive dynamic surface reconstructions on catalyst surfaces under reactive conditions[39–42]. These adsorbate-driven structural transformations underpin critical catalytic processes, including active site generation[3,4,43,44] and cyclic surface reconfigurations[39,45]. It is often speculated that reconstructions often precede the emergence of new surface phases in systems, such as oxide formation on transition metals catalysts like copper[39], however, direct evidence is generally not available. The present work addresses this important gap. Catalytic models that incorporate surface reconstructions and transient oxide phases are critical for simulating reaction pathways under operando conditions.

## 5. Acknowledgements

## 6. Author contributions

## 7. Supporting Information

*Commun.* **11**, 3554 (2020).